\documentclass[twocolumn,superscriptaddress,amsmath,amssymb,aps,pra]{revtex4}
\usepackage{amsmath,amsthm,amssymb}
\usepackage{latexsym}
\usepackage{amscd,graphicx}
\usepackage[titletoc]{appendix}
\usepackage{epsfig}
\usepackage{epstopdf}
\usepackage[normalem]{ulem}
\usepackage[dvipsnames]{xcolor}
\usepackage[colorlinks=true,linkcolor=red,citecolor=blue]{hyperref}

\begin{document}
	
	\newcommand{\cmt}[1]{{\textcolor{red}{#1}}}
	\newcommand{\rvs}[1]{{\textcolor{blue}{#1}}}

\title{Nonreciprocal Bistability in Coupled Nonlinear Cavity Magnonics}

\author{Wei Xiong}
\affiliation{Department of Physics, Wenzhou University, Zhejiang 325035, China}
\affiliation{International Quantum Academy, Shenzhen, 518048, China}

\author{Yuan Gong}
\affiliation{Department of Physics, Wenzhou University, Zhejiang 325035, China}

\author{Zhuanxia Li}
\affiliation{Department of Physics, Wenzhou University, Zhejiang 325035, China}

\author{Ying-Xia Wu}
\affiliation{Department of Physics, Wenzhou University, Zhejiang 325035, China}

\author{Yan-Xue Cheng}
\affiliation{Department of Physics, Wenzhou University, Zhejiang 325035, China}

\author{Jiaojiao Chen}
\altaffiliation{jjchenphys@wzu.edu.cn}
\affiliation{Department of Physics, Wenzhou University, Zhejiang 325035, China}

\date{\today }

\begin{abstract}
We propose a coupled nonlinear cavity-magnon system, consisting of two cavities, a second-order nonlinear element, and a yttrium-iron-garnet (YIG) sphere that supports Kerr magnons, to realize the sought-after highly tunable nonreciprocity. We first derive the critical condition for switching between reciprocity and nonreciprocity in the absence of magnon driving, and then numerically demonstrate that strong magnonic nonreciprocity can be achieved by violating this critical condition. When magnons are driven, we show that strong magnonic nonreciprocity can also be attained even within the critical condition. Compared to previous studies, the introduced nonlinear element not only relaxes the critical condition in both the weak and strong coupling regimes, but also offers an alternative means to tune magnonic nonreciprocity. Our work provides a promising avenue for realizing highly tunable nonreciprocal devices based on Kerr magnons.
\end{abstract}

\maketitle

\section{Introduction}
Thanks to the easily engineered strong coupling between photons and magnons~\cite{Huebl-2013,Tabuchi-2014,Zhang-2014,Goryachev-2014,Wang-2016,Bhoi-2014,Bai-2015,ZhangD-2015,Li-2019,Hou}, which refers to the collective spin excitations in ferro- and ferrimagnetic crystals such as yttrium iron garnet (YIG), a flourishing field of cavity magnonics has emerged and garnered significant attention in recent years~\cite{Rameshti-2022,Lachance-2019,Yuan-2022}. Experimentally, a sub-millimeter-scale YIG sphere coupled with a three-dimensional microwave cavity constitutes the most commonly employed cavity-magnon system~\cite{Tabuchi-2014,Zhang-2014,Goryachev-2014,Wang-2016}. With its flexible controllability and the long coherence time of magnons, cavity magnonics has become a fertile platform for investigating numerous exotic phenomena~\cite{Rameshti-2022,Wang-2020}. These include magnon memory~\cite{ZhangX-2015}, manipulation of spin currents~\cite{Bai-2015,Bai-2017,Mukhopadhyay-2022}, magnon entanglement~\cite{Yuanhy-2020,Mousolou-2021,ZhangZ-2019,Ren-2022,Pengjx2025,Ahmed-2025}, dissipative magnon–photon coupling~\cite{Harder-2018,Grigoryan-2018,Wang-2019}, magnon blockade~\cite{Huang-2018,Yao-2022,Wangy-2022,Xie-2020,Wang-2022}, non-Hermitian physics~\cite{ZhangD-2017,Harder-2017,Cao-2019,Zhao-2020,Zhanggq-2019}, cooperative polariton dynamics~\cite{Yao-2017}, enhancement of spin coupling~\cite{Tian-2023,Hei-2021}, quantum states of magnons~\cite{Xu-2023,Yuan-2020,Sun-2021,Zhanggq-2023,Qi-2022}, and microwave-to-optical transduction~\cite{Hisatomi-2016,Zhu-2020}. Additionally, research efforts have extended to exploring magnon-based hybrid quantum systems, such as qubit magnonics (with superconducting qubits or solid-state spins)~\cite{Tabuchi-2015,Lachance-2020,Dobrovolskiy-2019,Wolski-2020,Xiong4-2022,neuman-2020,neuman1-2021,neuman2-2021,Skogvoll-2021,trifunovic-2013,Fukami-2021}, cavity magnomechanics~\cite{zhang-2016,Li-2018,Shen-2022,Li-2021,Sohail-2025,Faisal-2025,Sohail-2023,Sohail-2023b}, cavity-magnon optomechanics~\cite{Chen,Proskurin-2018,Xiong-2023,Berinyuy2025,Sohail2023}, and cavity optomagnonics~\cite{Zhangx-2016,Osada-2016,Haigh-2016}.

With modern experimental techniques, the magnon Kerr effect, arising from the magnetocrystalline anisotropy in YIG~\cite{Zhangguoq-2019}, has been discovered and experimentally demonstrated~\cite{Wang-2016,Wangyp-2018} very recently. This has established the field of nonlinear cavity magnonics~\cite{Zheng-2023}, enabling the study of novel phenomena including bi- and tristability~\cite{Wang-2016,Shenrc-2021}, magnon–magnon entanglement~\cite{ZhangZ-2019}, strong spin–spin coupling~\cite{Xiong4-2022,Ji-2023,Skogvoll-2021,Peng2025}, superradiant phase transitions~\cite{Liu-2023}, and sensitive detection~\cite{Zhanggq1-2023,Nair-2021}. Moreover, the magnon Kerr effect can also be utilized to investigate nonreciprocal devices, such as nonreciprocal entanglement~\cite{Chen,Chen2024,Liu2025}, nonreciprocal transmission~\cite{Kong-2019}, nonreciprocal photon blockade~\cite{Fan2024}, nonreciprocal quantum synchronization~\cite{Lai2025}, and nonreciprocal higher-order sideband generation~\cite{Wangm-2021}, all of which are important and essential for quantum information processing and quantum networks. However, there is a growing demand for highly tunable nonreciprocal devices, and it still remains an open and challenging question in the field.

For this purpose, we propose a coupled nonlinear cavity–magnon system to realize highly tunable magnonic nonreciprocity by adjusting various system parameters. The system consists of a cavity embedded with a second-order nonlinear element coupled to another cavity housing a YIG sphere that supports Kerr magnons (i.e., magnons exhibiting Kerr nonlinearity). We first analytically establish the critical condition for switching between reciprocity and nonreciprocity in the absence of a magnon driving field. Specifically, when the condition is satisfied, the system exhibits magnonic reciprocity, whereas breaking this condition leads to magnonic nonreciprocity. {This arises because the Kerr nonlinearity effectively violates the Lorentz reciprocal theorem~\cite{Lorentz-1896,Masoud-2019}.} We then numerically demonstrate that highly tunable strong magnonic nonreciprocity can be achieved by tuning the photon–photon coupling strength, the coefficient of the nonlinear element, or both. Finally, we show that an external magnon driving field can induce strong magnonic nonreciprocity through fine-tuning of system parameters, even when the critical condition is unbroken. Compared to previous work~\cite{Kong-2019}, our results reveal that the introduced nonlinear element not only relaxes the critical condition across both weak- and strong-coupling regimes, but also provides an additional route for manipulating and achieving magnonic nonreciprocity. Our study thus offers a highly controllable platform for realizing nonreciprocal devices with Kerr magnons, promising versatile applications in nonlinear cavity magnonics.

The remainder of this paper is organized as follows. In Sec.~\ref{section2}, we describe the model and present the system’s effective Hamiltonian. In Sec.~\ref{section3}, we derive the critical condition that governs the transition between reciprocity and nonreciprocity in the absence of a magnon driving field. Section~\ref{section4} provides a numerical investigation of magnonic nonreciprocity under various system parameters by breaking the critical condition obtained in Sec.~\ref{section3}. Subsequently, in Sec.~\ref{section5}, we examine magnonic nonreciprocity in the presence of a magnon driving field while keeping the critical condition satisfied. Finally, conclusions are drawn in Sec.~\ref{section6}.

\section{Model and Hamiltonian}\label{section2}

  \begin{figure}
 	\includegraphics[scale=0.375]{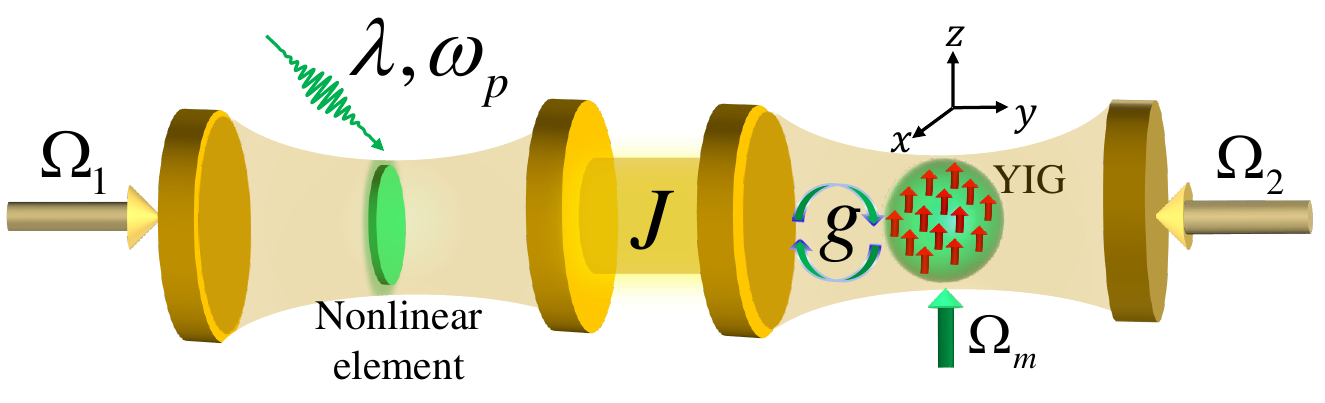}
 	\caption{Schematic diagram of the coupled nonlinear cavity magnonics. The system is composed of a PDC coupled to a MC embedded a YIG sphere with coupling strength $J$. The PDC can be realized by placing a second-order nonlinear element in a cavity. $g$ is the photon-magnon coupling strength, $\lambda$ is the parametric strength of the nonlinear element, $\omega_p$ is the frequency of the pumping field on the nonlinear element,  $\Omega_1$, $\Omega_2$, and $\Omega_m$ are three resonant driving fields. We here assume that the crystallographic axis of the YIG sphere is along the $z$-direction.}\label{fig1}
 \end{figure}
 
 We consider a coupled nonlinear cavity–magnon system consisting of a parametrically driven cavity (PDC) coupled to a magnonic cavity (MC) embedded with a micrometer-sized YIG sphere capable of supporting Kerr magnons (i.e., magnons exhibiting Kerr nonlinearity), as shown in Fig.~\ref{fig1}. The parametrically driven cavity can be realized by placing a pumped nonlinear element, such as a second-order nonlinear medium, inside the cavity~\cite{Qiu-2023,Mahboob-2022}. 
 {Experimentally, a YIG sphere with a diameter of approximately $1$ mm is typically employed, which is glued to the inner wall of a three-dimensional cavity at the magnetic-field antinode of the $\mathrm{TE}_{102}$ mode~\cite{Wang-2016}. By applying a static magnetic field generated by a superconducting magnet, the YIG sphere can be uniformly magnetized, thereby exciting the uniform Kittel mode with a low damping rate of about $1$ MHz~\cite{Huebl-2013}. The bias magnetic field is tunable within the range of $0$ to $1$ T~\cite{Wang-2016}, and its amplitude determines the eigenfrequency of the Kittel mode. To achieve a considerable Kerr nonlinearity, the applied magnetic field must be aligned along a particular crystallographic axis of the YIG~\cite{Zhanggq-2019}, which can be well controlled by adjusting the direction of the bias magnetic field~\cite{Wang-2016,Wangyp-2018}.}
 In addition, we assume that three driving fields with a common resonant frequency $\omega_d$ are applied respectively to the two cavities and to the YIG sphere. It should be noted that these external driving fields can be turned on and off at will. Considering the rotating-wave approximation, the Hamiltonian of the total system in the rotating frame with respect to frequency $\omega_d$ can be written as (setting $\hbar = 1$)
 
 \begin{align}\label{d1}
 	H=H_{\rm PDC}+H_{K}+H_{\rm CM}+H_{\rm PM}+H_{D},
 \end{align}
where
\begin{align}
	H_{\rm PDC}=\Delta_1 a_1^{\dagger} a_1+i\frac{\lambda}{2}(a_1^{\dag}a_1^{\dag}-a_1 a_1)
\end{align}
is the Hamiltonian of the PDC, with the frequency detuning $\Delta_1=\omega_1-\omega_d$ and the effective parametric strength $\lambda$ proportional to the Rabi frequency of the pumping field. Thus, both $\Delta_1$ and $\lambda$ are tuanble. $\omega_1$ is the eigenfrequency of the PDC and $a_1$ ($a_1^\dag$) is the annihilation (creation) operator of the PDC. {Experimentally, the PDC can be achieved by degenerate parametric down-conversion in flux-driven Josephson parametric amplifiers~\cite{Yurke-1987,Yurke-1988,Yurke-1989,Movshovich1990,Castellanos-Beltran2008,Yamamoto,Mallet2011,Fedorov2016,Kono2017,Bienfait2017}.}  The second term in Eq.~(\ref{d1}),
\begin{align}
	H_K = \Delta_{m}\, m^{\dagger} m + K\, m^{\dagger} m\, m^{\dagger} m,
\end{align}
denotes the Hamiltonian of Kerr magnons~\cite{ZhangZ-2019,Xiong4-2022}, with $m$ ($m^\dagger$) being the annihilation (creation) operator. Here, $\Delta_{m} = \omega_{m} - \omega_d$, with $\omega_m = \gamma H$, is the frequency detuning of the Kerr magnons from the driving field. The parameter $\gamma/2\pi = 28\,\mathrm{GHz/T}$ is the gyromagnetic ratio, and $H$ is the applied static magnetic field. The coefficient $K = \mu_{0} K_{\rm an} \gamma / (M^{2} V_{m})$ is the Kerr coefficient, where $\mu_{0}$ is the vacuum magnetic permeability, $K_{\rm an}$ is the first-order anisotropy constant, $M$ is the saturation magnetization, and $V_{m}$ is the volume of the YIG sphere. Experimentally, both the magnitude and sign of $K$ can be tuned by varying the direction of the bias magnetic field applied to the YIG sphere~\cite{Wangyp-2018}. For instance, the Kerr coefficient is positive (negative) when the bias field is aligned along the crystallographic axis $[100]$ ($[110]$), i.e., $K > 0$ ($K < 0$). Kerr magnons have been widely used to study bistability~\cite{Wangyp-2018} and tristability~\cite{Shenrc-2021}, nonreciprocity~\cite{Chen,Kong-2019}, sensitive detection~\cite{Zhanggq-2023}, quantum entanglement~\cite{Chen,ZhangZ-2019}, and quantum phase transitions~\cite{Liu-2023}. Since the magnitude $|K|$ is inversely proportional to the volume $V_m$, strong Kerr nonlinearity can be achieved by reducing the size of the YIG sphere down to the nanometer scale~\cite{Xiong4-2022}. The Hamiltonian $H_{\rm CM}$ in Eq.~(\ref{d1}) charaterizes the interaction between the Kerr magnons and MC, reading as
\begin{align}
	H_{\rm CM}=\Delta_2 a_2^{\dagger} a_2+g(a_2^\dagger m+a_2 m^\dagger),
\end{align}
where $\Delta_2=\omega_2-\omega_d$, with $\omega_2$ being the eigenfrequency of the MC, is the frequeny detuning of the MC from the driving field, and $g$ is the coupling strength between the Kerr magnons and MC with the annihilation (creation) operator $a_2$ ($a_2^\dag$). Experimentally, $g$ in the strong coupling regime has been demonstrated with sub-millimeter-sized YIG sphere and rich physics has been observed~\cite{Tabuchi-2014,Zhang-2014,Goryachev-2014,Wang-2016,ZhangD-2015}. The fourth term
\begin{align}
	H_{\rm PM}=J (a_1^\dag a_2+a_1 a_2^\dag)
\end{align}
characterizes the interaction between the PDC and MC with the coupling strength $J$. The last term $H_D$ in Eq.~(\ref{d1}) represents the field-matter coupling Hamiltonian, given by
\begin{align}
	H_D=i(\Omega_1 a_1^\dag+\Omega_2 a_2^\dag+\Omega_m m^\dag)+{\rm H.c.},\label{eq2}
\end{align}
where $\Omega_{1(2)}=\sqrt{\eta_{1(2)}\kappa_{1(2)}}\Omega_{1(2)}^{\rm in}$ with $\Omega_{1(2)}^{\rm in}=\sqrt{P_{1(2)}/\hbar\omega_d}$ is the amplitude of the driving field applied to the PDC (MC), $\Omega_m=\sqrt{\eta_m\gamma_m}\Omega_m^{\rm in}$ with $\Omega_m^{\rm in}=\sqrt{P_m/\hbar\omega_d}$ is the amplitude of the biased magnetic field applied to the Kerr magnons in the YIG sphere. Here, $\kappa_{1(2)}$ is the total loss rate of the PDC (MC), i.e., $\kappa_{1(2)}=\kappa_{1(2),0}+\kappa_{1(2),\rm ex}$, and $\gamma_{m}$ is the total loss rate of the Kerr magnons, where $\kappa_0$ is the intrinsic loss rate, $\kappa_{\rm ex}$ is the external loss, $\eta_{1(2)}$ denotes the dimensionless external loss rate over the total loss rate of the PDC (MC), i.e., $\eta_{1(2)}=\kappa_{1(2),e}/\kappa_{1(2)}$. $\eta_{m}$ is the dimensionless parameter for the Kerr magnons in the YIG sphere, $P_{1(2)}$ and $P_m$  are the input powers of the driving fields. Note that the degenerate two-photon generation condition is taken in Eq.~(\ref{d1}), i.e., $\omega_p=2\omega_d$.
		
\section{Nonreciprocal condition}\label{section3}

\subsection{Steady-state solution}
By taking dissipation into account, the dynamics of the system in Eq.~(\ref{d1}) can be given by the quantum Langevin equation,
\begin{align}\label{X1}
			\dot{a}_{1}&=-\left(i\Delta_{1}+\frac{\kappa_{1}}{2}\right)a_{1}-iJa_{2}+\lambda a^{\dagger}_{1}+\Omega_1,\notag\\
			\dot{a}_{2}&=-\left(i\Delta_{2}+\frac{\kappa_{2}}{2}\right)a_{2}-iJa_{1}-igm +\Omega_2,\\
		\dot{m}&=-\left[i(\Delta_{m}+2Km^{\dagger}m)+\frac{\gamma_m}{2}\right]m-iga_{2}+\Omega_m.\notag
\end{align}
In the long-time limit, the system can reach its steady state, resulting in the derivative of the expectation values ($a_{1s}$, $a_{2s}$, and $m_{s}$) for the system operators disappearing., i.e., $\dot{a}_{1s}=\dot{a}_{2s}=\dot{m_s}=0$. Thus, we have
\begin{align}\label{eq10}
	-\left(i\Delta_{1}+\frac{\kappa_{1}}{2}\right)a_{1s}-iJa_{2s}+\lambda a^*_{1s}+\Omega_1=&0,\notag\\
	-\left(i\Delta_{2}+\frac{\kappa_{2}}{2}\right)a_{2s}-iJa_{1s}-igm_s +\Omega_2=&0,\\
	-\left(i\tilde{\Delta}_m+\frac{\gamma_m}{2}\right)m_s-iga_{2s}+\Omega_m=&0,\notag
\end{align}
where $\tilde{\Delta}_m=\Delta_{m}+2K|m_s|^2$. From the third equation in  Eq.~(\ref{eq10}), $a_{2s}$ can be calculated directly, i.e.,
\begin{align}
	a_{2s}=\left[\Omega_m-(\gamma_m/2+i\tilde{\Delta}_m)m_s\right]/ig.
\end{align}
Then we insert $a_{2s}$ into the second equation in  Eq.~(\ref{eq10}), and we obtain
\begin{align}
	a_{1s}=&\left[(\kappa_2/2+i\Delta_2)\Omega_m-ig\Omega_2\right]/gJ\\
	&-\left[(\kappa_2/2+i\Delta_2)(\gamma_m/2+i\tilde{\Delta}_m) +g^2\right]m_s/gJ.\notag
\end{align}
Substituting $a_{2s}$, $a_{1s}$ and their conjugates into the first equation in Eq.~(\ref{eq10}), $m_s$ can be given by
\begin{align}
	m_s=\frac{A^*\Omega+B\Omega^*}{|B|^2-|A|^2},\label{eq13}
\end{align}
where 
\begin{align}
	A=&(\frac{\kappa_1}{2}+i\Delta_1)(\frac{\kappa_2}{2}+i\Delta_2)(\frac{\gamma_m}{2}+i\tilde{\Delta}_m)\notag\\
	&+g^2(\frac{\kappa_1}{2}+i\Delta_1)+J^2(\frac{\gamma_m}{2}+i\tilde{\Delta}_m),\notag\\
	B=&\lambda\left[(\frac{\kappa_2}{2}-i\Delta_2)(\frac{\gamma_m}{2}-i\tilde{\Delta}_m)+g^2\right],\\
	\Omega=&gJ\Omega_1+ig(\frac{\kappa_1}{2}+i\Delta_1+\lambda)\Omega_2\notag\\
	&-\left[(\frac{\kappa_1}{2}+i\Delta_1)(\frac{\kappa_2}{2}+i\Delta_2)-\lambda(\frac{\kappa_2}{2}-i\Delta_2)+J^2\right]\Omega_m.\notag
\end{align}
Obviously, there is a singularity for the magnon number $|m_s|^2\equiv M$, that is,  $|A|=|B|$. 
\subsection{Nonreciprocal condition}
{By taking the modulus of both sides of Eq.~(\ref{eq13}), one obtains a quintic polynomial equation that governs the steady-state magnon number \( M \), expressed as
	\begin{align}\label{eq15}
		c_5 M^5 + c_4 M^4 + c_3 M^3 + c_2 M^2 + c_1 M + c_0 = 0,
	\end{align}
	where the coefficients are given by
	\begin{align}\label{x12}
		c_5 &= d_2^2,
		c_4 = 2d_1 d_2,
		c_3 = d_1^2 + 2d_0 d_2, \\
		c_2 &= 2d_0 d_1 - |\gamma_1|^2,
		c_1 = d_0^2 - 2\,{\rm Re}(\gamma_0^* \gamma_1),
		c_0 = -|\gamma_0|^2,\notag
	\end{align}
	with $\chi_j=\kappa_j+i\Delta_j$ ($j=1,2,m$), $d_2 = |\beta_1|^2 - |\alpha_1|^2$, $d_1 = 2\,{\rm Re}(\beta_0^* \beta_1 - \alpha_0^* \alpha_1)$, $d_0 = |\beta_0|^2 - |\alpha_0|^2$. The relevant parameters are defined as
	\begin{align}\label{x13}
		\alpha_0 =& g^2 \chi_1 + (\chi_1 \chi_2 + J^2)\chi_m,~ \alpha_1 = 2iK(\chi_1 \chi_2 + J^2),\notag\\
		\beta_0 =& \lambda(\chi_2^* \chi_m^* + g^2), ~\beta_1 = -2iK\lambda \chi_2^*, \\
		\gamma_0 =& \alpha_0^* \Omega + \beta_0 \Omega^*, ~\gamma_1 = \alpha_1^* \Omega + \beta_1 \Omega^*.\notag
	\end{align}
	Since our interest lies in the system's response to cavity input fields, we set $\Omega_m = 0$ for simplicity. Under this condition, it follows from Eqs.~(\ref{x12}) and~(\ref{x13}) that the coefficients $c_2$, $c_1$, and $c_0$ are only dependence of the input field on the cavity. Specifically, in the presence of the left (right) input field, i.e., \(\Omega_1 \neq 0\) but \(\Omega_2 = 0\) (\(\Omega_2 \neq 0\) but \(\Omega_1 = 0\)), we have $c_2=c_2[\Omega_{1(2)}], c_1=c_1[\Omega_{1(2)}]$, and $c_0=c_0[\Omega_{1(2)}]$. When the following impedance-matching condition
	\begin{align}\label{X8}
		c_2[\Omega_1] = c_2[\Omega_2],~
		c_1[\Omega_1] = c_1[\Omega_2],~
		c_0[\Omega_1] = c_0[\Omega_2],
\end{align}
is satisfied, the mean magnon number $M$ takes the same value for the left or right input field. Therefore, the emergence of asymmetric magnonic responses is directly associated with the breaking of this impedance-matching condition.

By solving Eq.~(\ref{X8}), the impedance-matching condition can be explicitly written as}
\begin{align}\label{eq19}
	J_c^{(0)}=&\frac{\Omega_2 }{\Omega _1}\sqrt{\Delta _1^2+\kappa _1^2/4},~\lambda_c^{(0)}= 0,
\end{align}
or
\begin{align}\label{eq20}
	J_c=\frac{ \Omega_2}{\Omega _1}\Delta _1,~\lambda_c= -\frac{\kappa _1}{2}.
\end{align}
Eq.~(\ref{eq19}) implies that the magnon number can become nonreciprocal when $J \neq J_c^{(0)}$ even in the absence of the nonlinear element ($\lambda = \lambda_c^{(0)} = 0$), a situation that has been previously investigated~\cite{Kong-2019} and will not be discussed further here. When the nonlinear element ($\lambda \neq 0$) is introduced, the critical parameters are modified to $J = J_c$ and $\lambda = \lambda_c$, as given by Eq.~(\ref{eq20}). This means that the newly added nonlinear element provides an adjustable means to induce magnonic nonreciprocity even when $J = J_c$ is fixed. In fact, the coupling strength $J$ between the two cavities can also be tuned experimentally. Therefore, the magnonic nonreciprocity in our proposal can be achieved by tuning both $J$ and $\lambda$, offering significantly enhanced controllability compared with the previous scheme~\cite{Kong-2019}. Moreover, the introduced nonlinear element can substantially {\it relax} the critical condition to both the weak- and strong-coupling regimes, because $J_c^{(0)} \propto \sqrt{\Delta_1^{\,2} + \kappa_1^{\,2}/4} > J_c \propto \Delta_1$. This indicates that a reciprocal magnon number can be realized in both the strong- and weak-coupling regimes when the nonlinear element is included, whereas in the previous study~\cite{Kong-2019}, reciprocity could be achieved only in the strong-coupling regime.

\begin{figure}
	\includegraphics[scale=0.45]{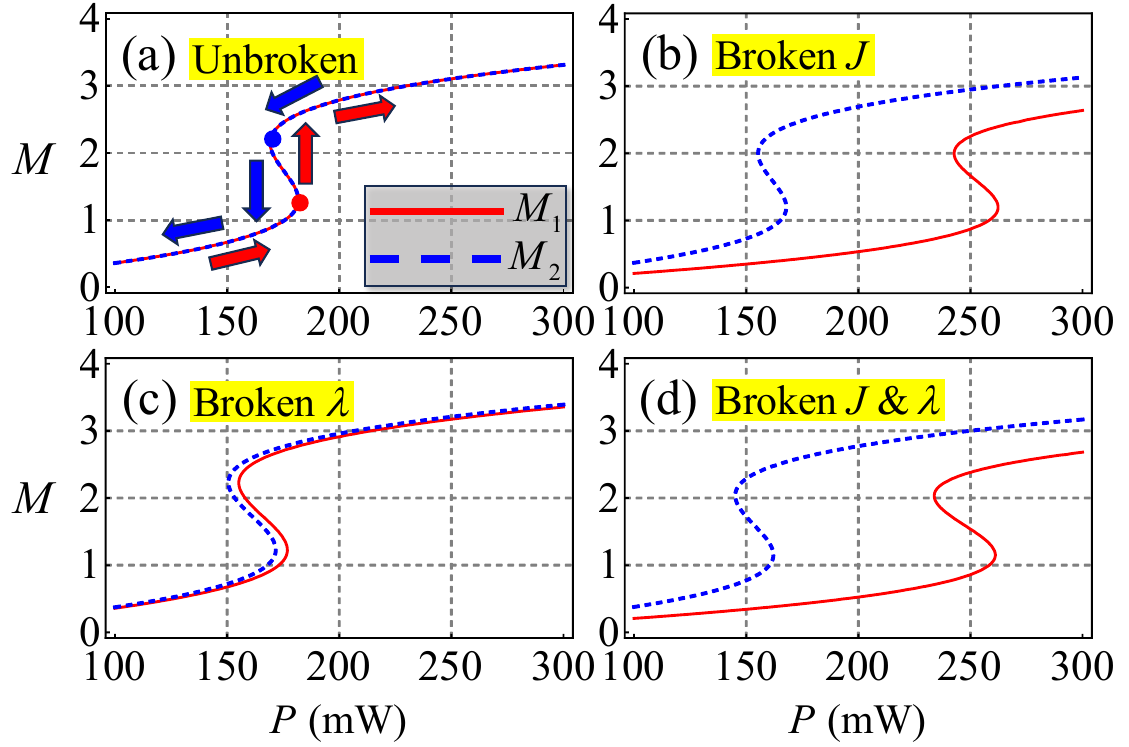}
	\caption{The scaled magnon number as a function of the input power ($P_1=P_2=P$) of the driving field on the PDC (MC) with the critical condition in Eq.~(\ref{eq20}) (a) unbroken and (b, c, d) broken, where (a) $J=J_c$ and $\lambda=\lambda_c$, (b) $J=0.8J_c$ and $\lambda=\lambda_c$, (c) $J=J_c$ and $\lambda=0.2\lambda_c$, and (d) $J=0.8J_c$ and $\lambda=0.2\lambda_c$. {The red solid (blue dashed) curve denotes the magnon number when the PDC (MC) is driven.} Other parameters are chosen as: $\eta_1=\eta_2=\eta_m=0.5$, $\omega_m/2\pi=\omega_d/2\pi=10.1$ GHz, $g/2\pi=41$ MHz, $\gamma_m/2\pi=20$ MHz, $\kappa_1/2\pi=5\kappa_2/2\pi=25$  MHz, $\Delta_1=\Delta_2=4\gamma_m$, $\Delta_m=\omega_m-\omega_d$, $K/2\pi=0.5$ $\mu$Hz, and $P_m=0$.}\label{fig2}
\end{figure}

\section{Magnonic nonreciprocity without magnon driving}\label{section4}

Below, we investigate how the magnon number responds to system parameters when the critical condition in Eq.~(\ref{eq20}) is either satisfied or violated. In Fig.~\ref{fig2}, we plot the scaled magnon number as a function of the input power of the driving field on the PDC or MC, taking $P_1=P_2=P$. The magnon number clearly exhibits bistability. When the input power is increased, the magnon number follows the lower stable branch until it reaches a turning point [red point in Fig.~\ref{fig2}(a)], after which it jumps to the upper stable branch, as indicated by the red arrow. When the input power is subsequently decreased, the system evolves along the upper stable branch until it reaches the other turning point [blue point in Fig.~\ref{fig2}(a)], then switches back to the lower stable branch [blue arrow] and continues to decrease. Numerically, we find that the bistable magnon response is reciprocal when the critical condition in Eq.~(\ref{eq20}) is unbroken. However, once this condition is violated, either by setting $J\neq J_c$ or $\lambda\neq \lambda_c$, the magnon number becomes nonreciprocal [Figs.~\ref{fig2}(b-d)]. Furthermore, the degree of nonreciprocity is enhanced more effectively by breaking the coupling strength $J$ than by breaking the parametric strength $\lambda$ [compare Figs.~\ref{fig2}(b) and \ref{fig2}(c)], and it becomes even stronger when both $J$ and $\lambda$ are simultaneously detuned [Fig.~\ref{fig2}(d)]. {These results indicate that the observed nonreciprocal bistability originates from the violation of the reciprocity condition in Eq.~(\ref{eq20}), rather than from exceptional points in non-Hermitian systems.}

\begin{figure}
	\includegraphics[scale=0.452]{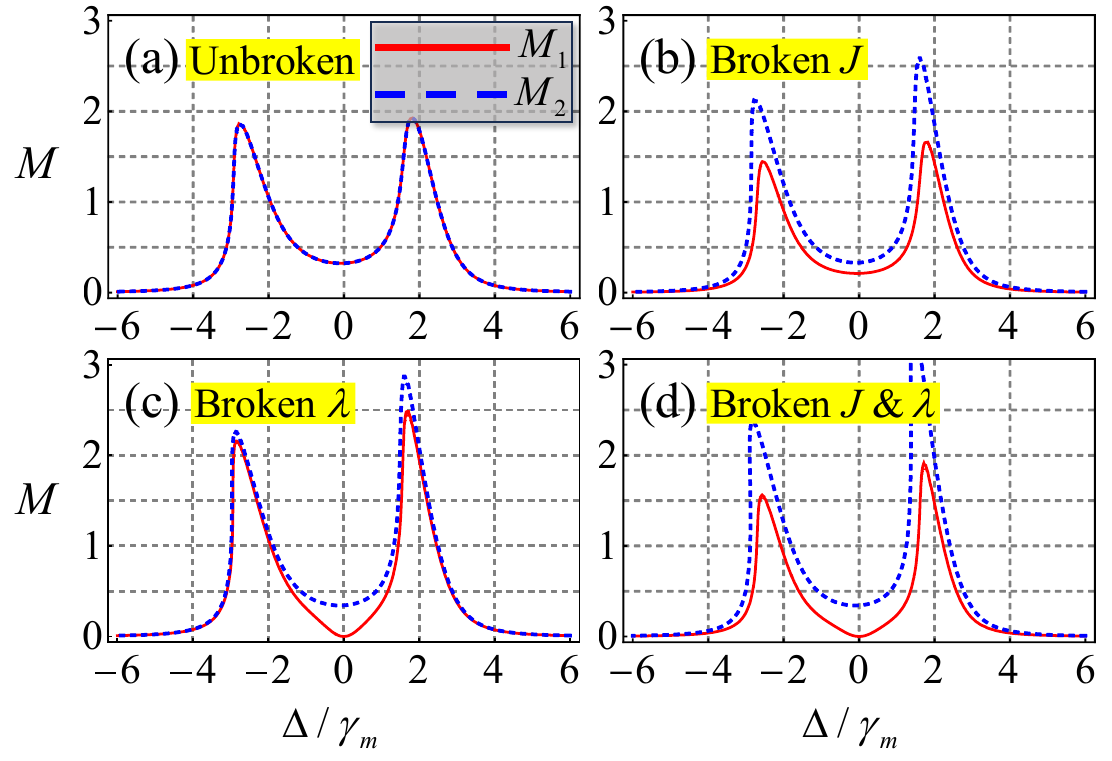}
	\caption{The scaled magnon number as a function of the frequency detuning ($\Delta_1=\Delta_2=\Delta$) of the PDC (MC) from the driving field with the critical condition in Eq.~(\ref{eq20}) (a) unbroken and (b,c,d) broken, where (a) $J=J_c$ and $\lambda=\lambda_c$, (b) $J=0.8J_c$ and $\lambda=\lambda_c$, (c) $J=J_c$ and $\lambda=0.2\lambda_c$, and (d) $J=0.8J_c$ and $\lambda=0.2\lambda_c$. {The red solid (blue dashed) curve denotes the magnon number when the PDC (MC) is driven.} Other parameters are the same as those in Fig.~\ref{fig2} except for $P_1=P_2=P=100$ mW, $\Delta_m=\Delta_1=\Delta_2=\Delta$, and $\omega_d=\omega_m-\Delta$.}\label{fig3}
\end{figure}

Figure \ref{fig3} shows the behavior of the scaled magnon number as a function of the frequency detuning $\Delta$ of the PDC or MC when the critical condition in Eq.~(\ref{eq20}) is either satisfied or violated, with $\Delta_1 = \Delta_2 = \Delta$ and $P_1 = P_2 = 100$~mW. When the critical condition is unbroken, the magnon number behaves reciprocally with respect to the driving field on the PDC or MC [see Fig.~\ref{fig3}(a)], as predicted theoretically. {By varying $\Delta$, the magnon number exhibits a two-peak profile [Fig.~\ref{fig3}(a)]. This occurs because $\Delta$ is chosen to be comparable to the coupling strength $J$ [see Eq.~(\ref{eq20})]. As $\Delta$ increases, the effective coupling $J$ becomes stronger, leading to the formation of two polaritons through the hybridization of the PDC and MC modes. The positions of the two peaks correspond to the eigenfrequencies of these polaritons.} When the critical condition is broken by setting $J \neq J_c$ or $\lambda \neq \lambda_c$, the magnon number not only retains the two-peak structure but also exhibits nonreciprocity [Figs.~\ref{fig3}(b--d)]. When only the coupling strength is detuned ($J \neq J_c$ while $\lambda = \lambda_c$), the parameter range of $\Delta$ over which magnonic nonreciprocity is observed is broader than in the case of detuning the parametric strength alone ($\lambda \neq \lambda_c$ while $J = J_c$), as shown by comparing Figs.~\ref{fig3}(b) and \ref{fig3}(c). Moreover, the nonreciprocity induced by breaking $J$ is significantly stronger at the two peaks than that caused by breaking $\lambda$, although near resonance ($\Delta \approx 0$), the trend reverses. The cooperative effect of simultaneously detuning both $J$ and $\lambda$ yields the strongest nonreciprocity over the widest parameter range [see Fig.~\ref{fig3}(d)].

\begin{figure}
	\includegraphics[scale=0.45]{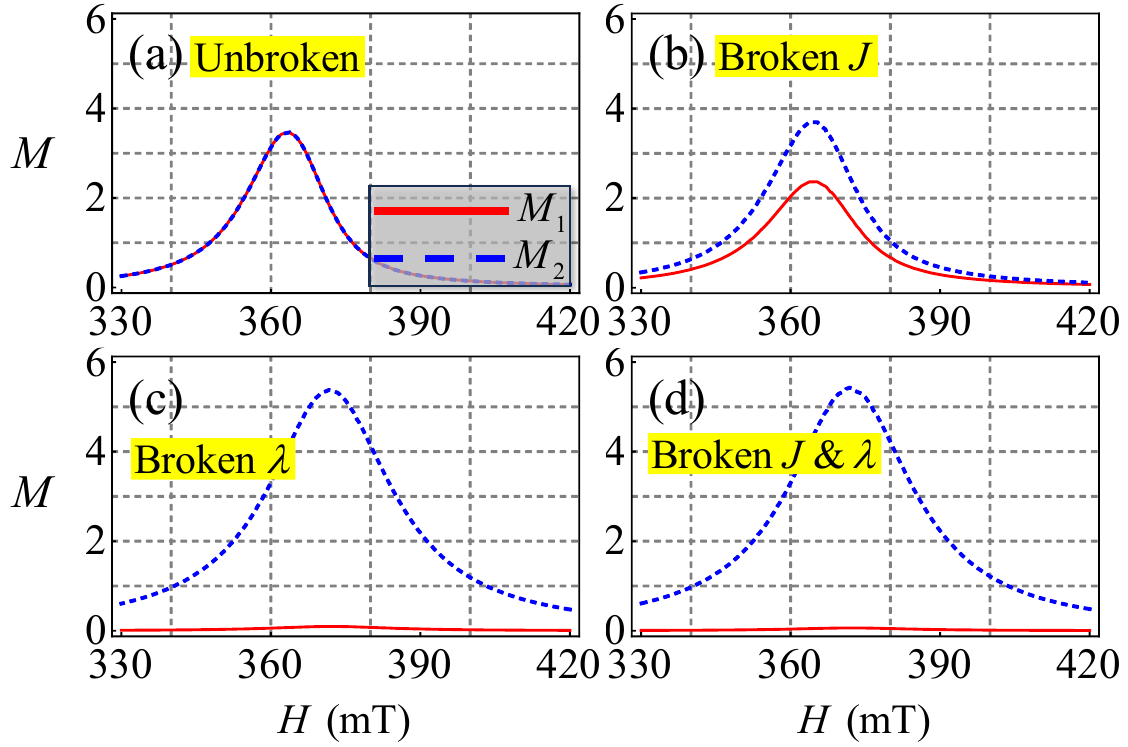}
	\caption{The scaled magnon number as a function of the biased magnetic field ($H=\omega_m/\gamma$) with the critical condition in Eq.~(\ref{eq20}) (a) unbroken and (b,c,d) broken, where (a) $J=J_c$ and $\lambda=\lambda_c$, (b) $J=0.8J_c$ and $\lambda=\lambda_c$, (c) $J=J_c$ and $\lambda=0.2\lambda_c$, and (d) $J=0.8J_c$ and $\lambda=0.2\lambda_c$. {The red solid (blue dashed) curve denotes the magnon number when the PDC (MC) is driven.} Other parameters are the same as those in Fig.~\ref{fig2} except for $P_1=P_2=P=100$ mW and $\Delta_1=\Delta_2=\Delta=0.1\gamma_m$.}\label{fig4}
\end{figure}
We then study the behavior of the magnon number as a function of the bias magnetic field $H$ when the critical condition in Eq.~(\ref{eq20}) is either satisfied or violated, as shown in Fig.~\ref{fig4}. When the critical condition is unbroken, the magnon number is reciprocal and exhibits a Lorentzian profile with respect to $H$ for the driving field on either the PDC or the MC [see Fig.~\ref{fig4}(a)]. However, when the critical condition is broken, the magnon number exhibits nonreciprocal behavior [Figs.~\ref{fig4}(b-d)]. Figure~\ref{fig4}(b) shows that, by breaking $J$ while keeping $\lambda = \lambda_c$, the magnon number induced by the driving field on the PDC is slightly suppressed around $H = 370$~mT (blue dashed curve), whereas the magnon number induced by the driving field on the MC is slightly enhanced (red solid curve), resulting in visible nonreciprocity. In Fig.~\ref{fig4}(c), breaking $\lambda = \lambda_c$ leads to much stronger nonreciprocity, with the magnon number driven by the PDC greatly enhanced and that driven by the MC nearly fully suppressed. When both $J = J_c$ and $\lambda = \lambda_c$ are simultaneously broken, the resulting nonreciprocity is dominated by the effect of breaking $\lambda = \lambda_c$ [Fig.~\ref{fig4}(d)], and the contribution from breaking $J = J_c$ can be neglected.

\begin{figure}
	\includegraphics[scale=0.45]{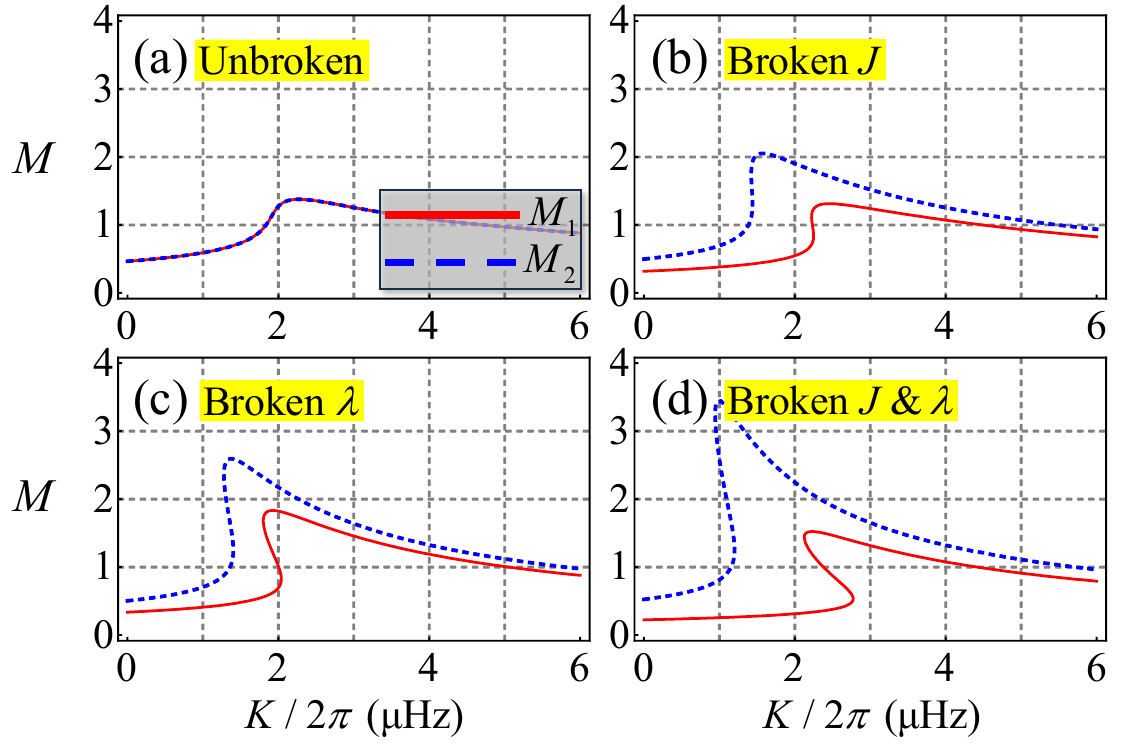}
	\caption{The scaled magnon number as a function of the Kerr coefficient ($K$) with the critical condition in Eq.~(\ref{eq20}) (a) unbroken and (b,c,d) broken, where (a) $J=J_c$ and $\lambda=\lambda_c$, (b) $J=0.8J_c$ and $\lambda=\lambda_c$, (c) $J=J_c$ and $\lambda=0.2\lambda_c$, and (d) $J=0.8J_c$ and $\lambda=0.2\lambda_c$. {The red solid (blue dashed) curve denotes the magnon number when the PDC (MC) is driven.} Other parameters are the same as those in Fig.~\ref{fig2} except for $P_1=P_2=P=100$ mW and $\Delta_1=\Delta_2=\Delta=\gamma_m$.}\label{fig5}
\end{figure}
Finally, we numerically examine the impact of the Kerr coefficient $K$ on the scaled magnon number when the critical condition in Eq.~(\ref{eq20}) is either satisfied or violated, as illustrated in Fig.~\ref{fig5}. When the critical condition is unbroken, the magnon number responds reciprocally to the driving field on the PDC or MC as $K$ is varied, producing an $S$-shaped profile [see Fig.~\ref{fig5}(a)]. However, when the critical condition is broken by detuning $J$ ($J \neq J_c$) [Fig.~\ref{fig5}(b)], the magnon number responds nonreciprocally to the driving field on the PDC or MC. In particular, the magnon number is enhanced (suppressed) and exhibits a redshift (blueshift) for $\Omega_1 \neq 0~(\Omega_2 \neq 0)$, giving rise to a sub-$S$ profile. When the critical condition is broken by detuning $\lambda$ (e.g., $\lambda = 0.2 \lambda_c$), the magnon numbers corresponding to $\Omega_1 = 0$ or $\Omega_2 = 0$ are nonreciprocally enhanced and redshifted, resulting in a genuine $S$-shaped profile [Fig.~\ref{fig5}(c)]. By simultaneously breaking both $J = J_c$ and $\lambda = \lambda_c$, the magnon number driven by the PDC is significantly enhanced and redshifted, while that driven by the MC is slightly increased and blueshifted, as shown in Fig.~\ref{fig5}(d). Moreover, this cooperative effect can produce a standard $S$-shaped bistable profile for the case of $\Omega_2 \neq 0$ [red curve in Fig.~\ref{fig5}(d)].

\section{Magnonic nonreciprocity with magnon driving}\label{section5}

\begin{figure}
	\includegraphics[scale=0.45]{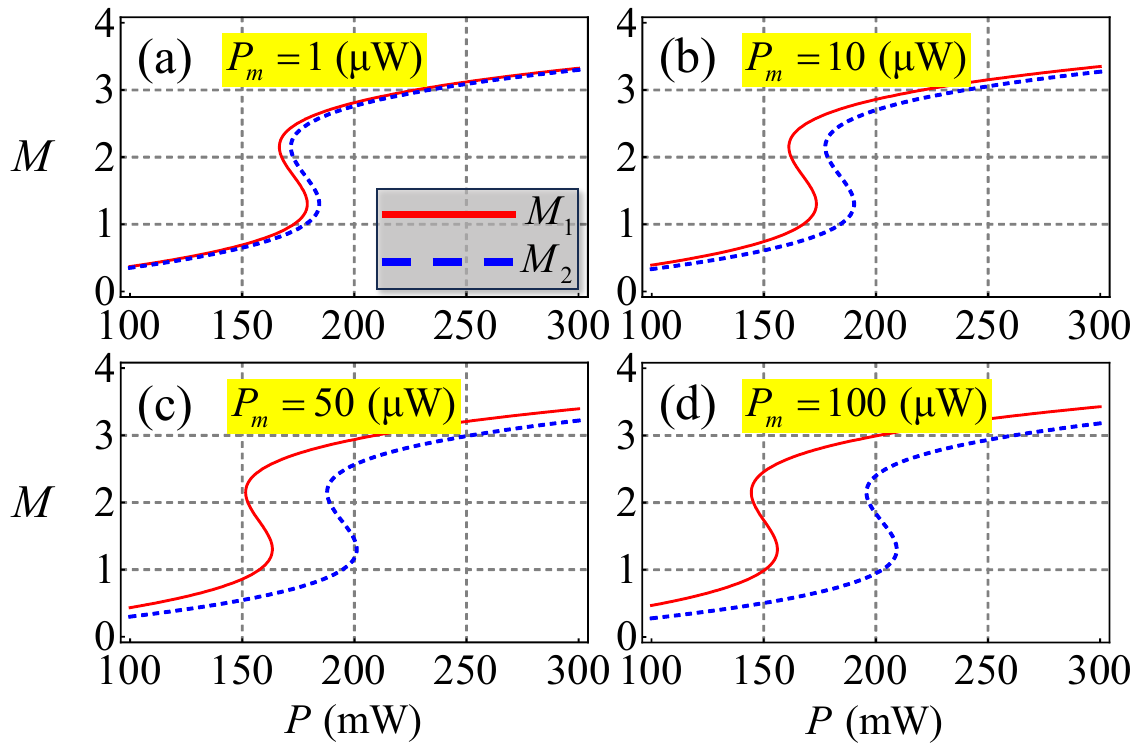}
	\caption{The scaled magnon number as a function of the input power ($P_1=P_2=P$) of the driving field on the PDC (MC) with the magnon driving field and the critical condition in Eq.~(\ref{eq20}) unbroken. In (a) $P_m=1$ $\mu$W, (b) $P_m=10$ $\mu$W, (c) $P_m=50$ $\mu$W, and (d) $P_m$=100 $\mu$W. {The red solid (blue dashed) curve denotes the magnon number when the PDC (MC) is driven.} Other parameters the same as those in Fig.~\ref{fig2}.}\label{fig6}
\end{figure}
\begin{figure}
\includegraphics[scale=0.45]{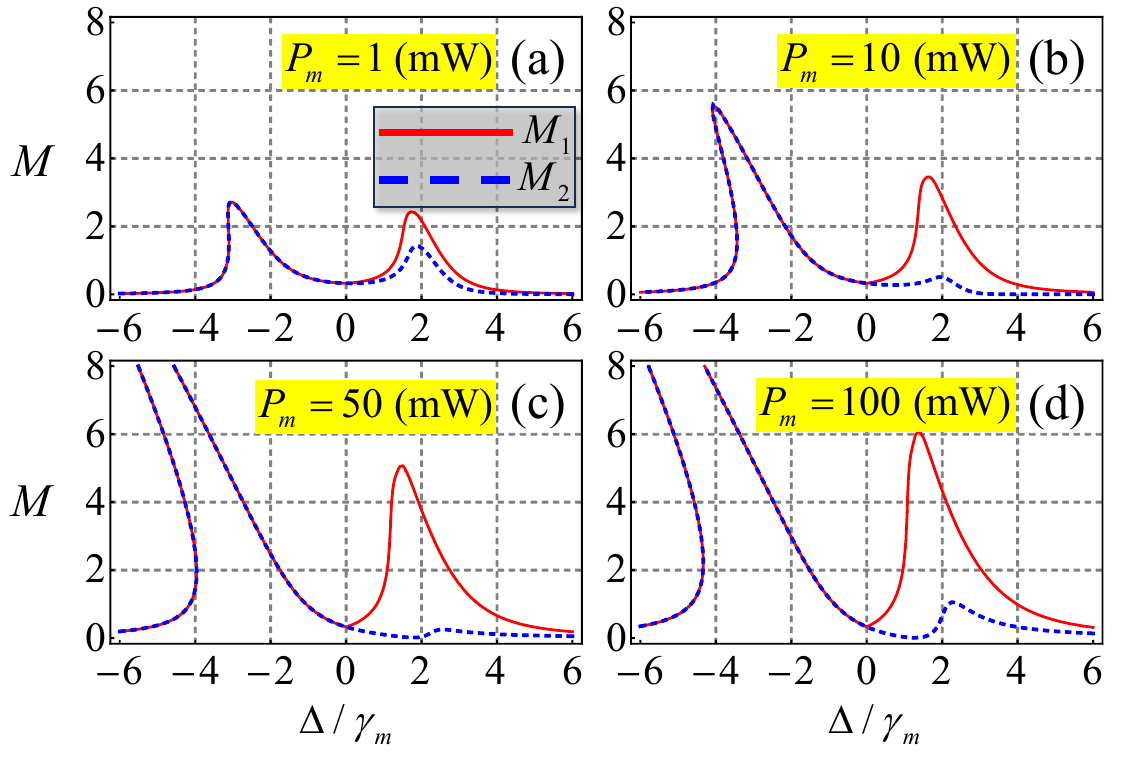}
\caption{The scaled magnon number as a function of the frequency detuning ($\Delta_1=\Delta_2=\Delta$) of the PDC (MC) from the driving field in the presence of the magnon driving field with the critical condition in Eq.~(\ref{eq20}) unbroken. In (a) $P_m=1$ mW, (b) $P_m=10$ mW, (c) $P_m=50$ mW, and (d) $P_m$=100 mW. {The red solid (blue dashed) curve denotes the magnon number when the PDC (MC) is driven.} Other parameters the same as those in Fig.~\ref{fig3}.}\label{fig7}
\end{figure}
\begin{figure}
	\includegraphics[scale=0.45]{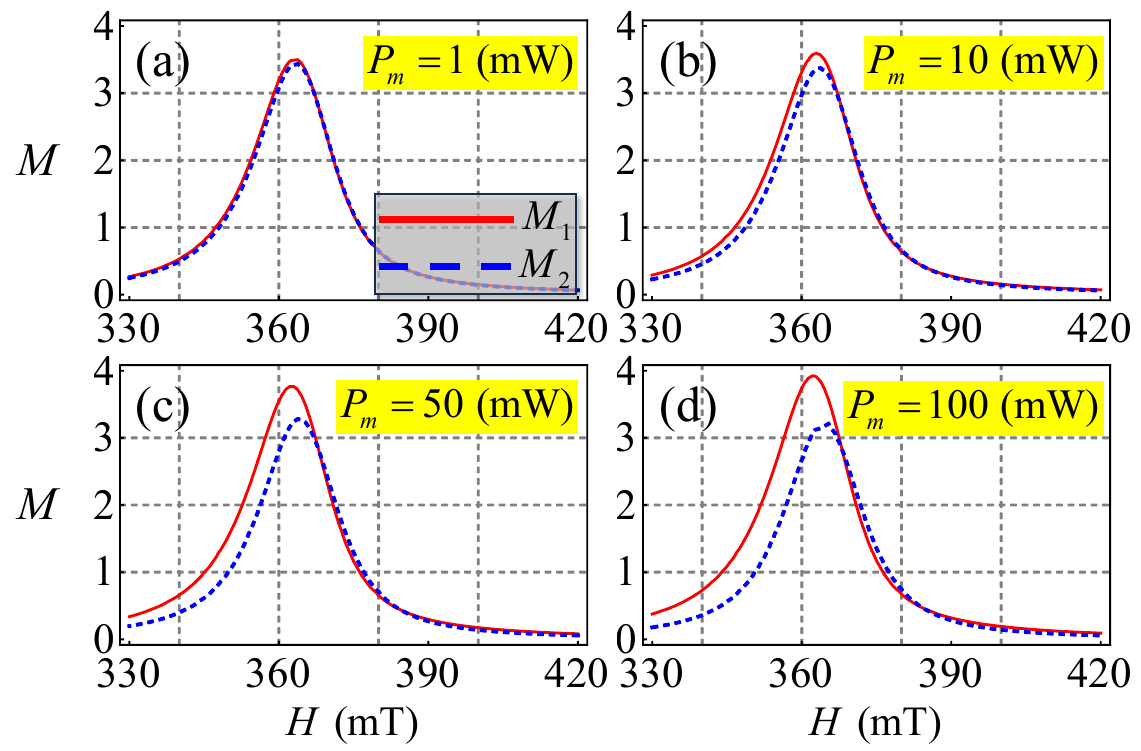}
	\caption{The scaled magnon number as a function of the biased magnetic field ($H=\omega_m/\gamma$) in the presence of the magnon driving field with the critical condition in Eq.~(\ref{eq20}) unbroken. In (a) $P_m=1$ mW, (b) $P_m=10$ mW, (c) $P_m=50$ mW, and (d) $P_m$=100 mW. {The red solid (blue dashed) curve denotes the magnon number when the PDC (MC) is driven.} Other parameters the same as those in Fig.~\ref{fig4}.}\label{fig8}
\end{figure}
\begin{figure}
	\includegraphics[scale=0.45]{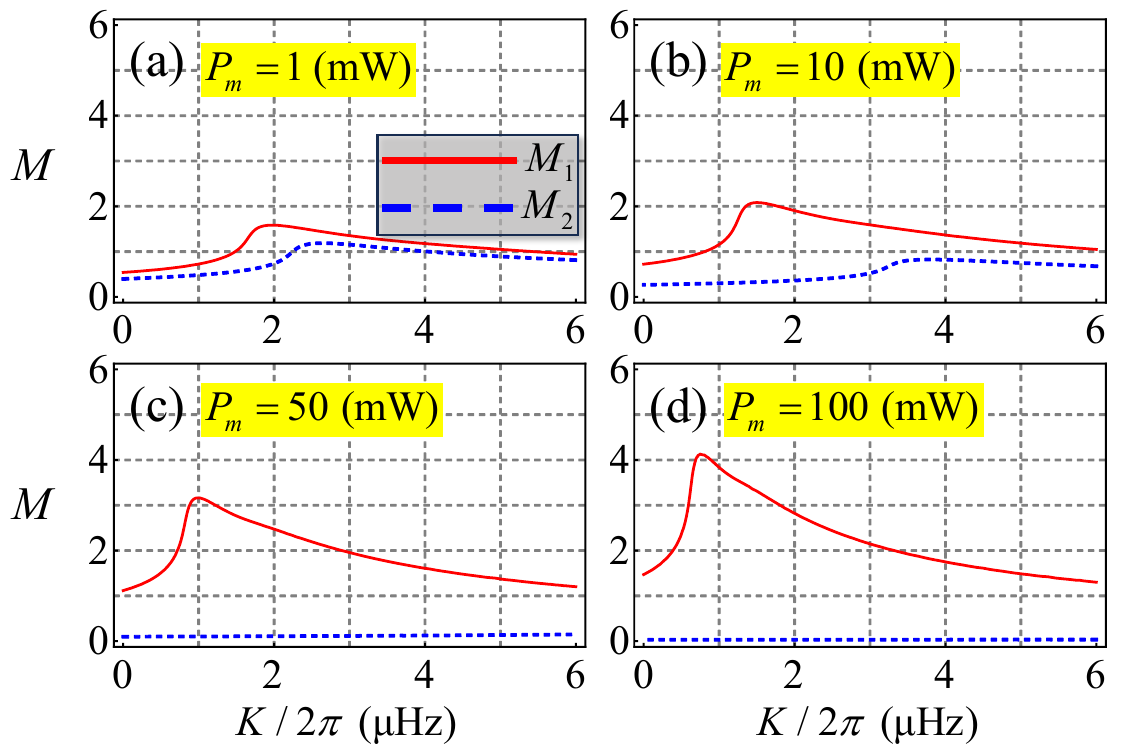}
	\caption{The scaled magnon number as a function of the Kerr coefficient ($K$) in the presence of the magnon driving field with the critical condition in Eq.~(\ref{eq20}) unbroken. In (a) $P_m=1$ mW, (b) $P_m=10$ mW, (c) $P_m=50$ mW, and (d) $P_m$=100 mW. {The red solid (blue dashed) curve denotes the magnon number when the PDC (MC) is driven.} Other parameters the same as those in Fig.~\ref{fig5}.}\label{fig9}
\end{figure}

It should be noted that the critical condition in Eq.~(\ref{eq20}) is derived for $\Omega_m = 0$, i.e., when the magnons in the YIG sphere are not driven. When a magnon driving field is applied ($\Omega_m \neq 0$), the condition in Eq.~(\ref{eq20}) is naturally altered. This implies that magnonic nonreciprocity can be induced by introducing an additional driving field to the magnons.

To elucidate the impact of the magnon driving field on the magnon number, we illustrate its dependence on various system parameters for different driving powers $P_m$ in Figs.~\ref{fig6}-\ref{fig9}. Figure~\ref{fig6} shows that the magnon number exhibits nonreciprocal bistability in the presence of the magnon driving field. For weak driving ($P_m = 1$~mW), a weak nonreciprocal bistability can be observed [Fig.~\ref{fig6}(a)]. As the power of the magnon driving field increases, the nonreciprocity becomes more pronounced [Figs.~\ref{fig6}(b)-\ref{fig6}(d)], and the range of input power over which nonreciprocity is observed is broadened. Figure~\ref{fig7} shows the magnon number as a function of frequency detuning $\Delta$. In the presence of the magnon driving field ($P_m \neq 0$), the magnon number behaves nonreciprocally in the red-detuned regime ($\Delta > 0$) but remains reciprocal in the blue-detuned regime ($\Delta < 0$). Moreover, stronger magnon driving powers further enhance the nonreciprocity by boosting (suppressing) the magnon number induced by the driving field on the MC (PDC). In Fig.~\ref{fig8}, we plot the magnon number as a function of the magnetic field $H$ for different magnon driving powers. Noticeably, significant nonreciprocity is observed only in the region $H < 365$~mT when assisted by a strong driving field, e.g., $P_m = 50$~mW [Fig.~\ref{fig8}(c)] or $P_m = 100$~mW [Fig.~\ref{fig8}(d)], while the magnon number behaves reciprocally in the opposite region. Finally, Fig.~\ref{fig9} depicts the magnon number as a function of the Kerr coefficient in the presence of the magnon driving field. It is clear that, with increasing $P_m$, the magnon number driven by the PDC (MC) is enhanced (suppressed) and experiences a significant redshift (blueshift). These results indicate that a strong nonreciprocity can be effectively realized by employing a sufficiently powerful magnon driving field.

\section{Conclusion}\label{section6}
In summary, we have proposed a coupled nonlinear cavity–magnon system, consisting of a cavity embedded with a nonlinear element coupled to another cavity containing Kerr magnons in a YIG sphere, to realize magnonic nonreciprocity. We first analytically derived the critical condition for predicting magnonic nonreciprocity in the absence of a magnon driving field. We then numerically demonstrated that strong magnonic nonreciprocity can be achieved by breaking the critical condition, through tuning the two-cavity coupling strength $J$, the parametric coupling strength $\lambda$, or both. Furthermore, we investigated magnonic nonreciprocity in the presence of a magnon driving field while maintaining the critical condition. The results show that a sufficiently strong magnon driving field can induce pronounced nonreciprocity within a specific parameter space. Compared to previous studies without the nonlinear element~\cite{Kong-2019}, our analysis reveals that the introduced nonlinear element not only relaxes the critical condition in both the weak- and strong-coupling regimes, thereby improving experimental feasibility, but also provides an additional pathway to manipulate magnonic nonreciprocity. Our study thus offers a promising platform for realizing highly tunable nonreciprocal devices based on Kerr magnons.

This work was supported by the Zhejiang Province Key R\&D Program of China (Grant No.~2025C01028), the Natural Science Foundation of Zhejiang Province (Grant No.~LY24A040004),  and the Shenzhen International Quantum Academy (Grant No.~SIQA2024KFKT010).


\begin{thebibliography}{99} 
\bibitem{Huebl-2013}H. Huebl, C. W. Zollitsch, J. Lotze, F. Hocke, M. Greifenstein, A. Marx, R. Gross, and S. T. B. Goennenwein, High Cooperativity in Coupled Microwave Resonator Ferrimagnetic Insulator Hybrids, Phys. Rev. Lett. {\bf 111}, 127003 (2013).

\bibitem{Tabuchi-2014}Y. Tabuchi, S. Ishino, T. Ishikawa, R. Yamazaki, K. Usami, and Y. Nakamura, Hybridizing Ferromagnetic Magnons and Microwave Photons in the Quantum Limit, Phys. Rev. Lett. {\bf 113}, 083603 (2014).

\bibitem{Zhang-2014}X. Zhang, C.-L. Zou, L. Jiang, and H. X. Tang, Strongly Coupled Magnons and Cavity Microwave Photons, Phys. Rev. Lett. {\bf 113}, 156401 (2014).

\bibitem{Goryachev-2014} M. Goryachev, W. G. Farr, D. L. Creedon, Y. Fan, M. Kostylev, and M. E. Tobar, High-Cooperativity Cavity QED with Magnons at Microwave Frequencies, Phys. Rev. Applied {\bf 2}, 054002 (2014).

\bibitem{Wang-2016}Y. P. Wang, G. Q. Zhang, D. Zhang, X. Q. Luo, W. Xiong, S. P.
Wang, T. F. Li, C. M. Hu, and J. Q. You, Magnon Kerr effect in a strongly coupled cavity-magnon system, Phys. Rev. B {\bf 94}, 224410 (2016).

\bibitem{Bhoi-2014}B. Bhoi, T. Cliff, I. S. Maksymov, M. Kostylev, R. Aiyar,
N. Venkataramani, S. Prasad, and R. L. Stamps, Study of photon-magnon coupling in a YIG-film split-ring resonant system, J. Appl. Phys. {\bf 116}, 243906 (2014).

\bibitem{Bai-2015}L. Bai, M. Harder, Y. P. Chen, X. Fan, J. Q. Xiao, and
C.-M. Hu, Spin Pumping in Electrodynamically Coupled Magnon-Photon Systems, Phys. Rev. Lett. {\bf 114}, 227201 (2015).

\bibitem{ZhangD-2015}D. Zhang, X. M. Wang, T. F. Li, X. Q. Luo, W. Wu,
F. Nori, and J. Q. You, Cavity quantum electrodynamics with ferromagnetic magnons in a small yttrium-iron-garnet sphere, npj Quantum Inf. {\bf 1}, 15014
(2015).

\bibitem{Li-2019}Y. Li, T. Polakovic, Y.-L. Wang, J. Xu, S. Lendinez, Z. Zhang, J. Ding, T. Khaire, H. Saglam, R. Divan, J. Pearson, W.-K. Kwok, Z. Xiao, V. Novosad, A. Hoffmann, and W. Zhang, Strong Coupling between Magnons and Microwave Photons in On-Chip Ferromagnet-Superconductor Thin-Film Devices, Phys. Rev. Lett. {\bf 123}, 107701 (2019).

\bibitem{Hou}J. T. Hou and L. Liu, Strong Coupling between Microwave Photons and Nanomagnet Magnons, Phys. Rev. Lett. {\bf 123}, 107702 (2019).

\bibitem{Rameshti-2022}B. Z. Rameshti, S. V. Kusminskiy, J. A. Haigh, K. Usami,
D. Lachance-Quirion, Y. Nakamura, C. M. Hu, H. X. Tang, G. E. W. Bauer, and Y. M. Blanter, Cavity magnonics, Phys. Rep. {\bf 979}, 1 (2022).

\bibitem{Lachance-2019}D. Lachance-Quirion, Y. Tabuchi, A. Gloppe, K. Usami, and
Y. Nakamura, Hybrid quantum systems based on magnonics,
Appl. Phys. Express {\bf 12}, 070101 (2019).

\bibitem{Yuan-2022}H. Y. Yuan, Y. Cao, A. Kamra, R. A. Duine, and P. Yan,
Quantum magnonics: When magnon spintronics meets quantum information science, Phys. Rep. {\bf 965}, 1 (2022).

\bibitem{Wang-2020}Y. P. Wang and C.-M. Hu, Dissipative couplings in cavity
magnonics, J. Appl. Phys. {\bf 127}, 130901 (2020).

\bibitem{ZhangX-2015}X. Zhang, C.-L. Zou, N. Zhu, F. Marquardt, L. Jiang, and
H. X. Tang, Magnon dark modes and gradient memory, Nat. Commun. {\bf 6}, 8914 (2015).

\bibitem{Bai-2017}L. Bai, M. Harder, P. Hyde, Z. Zhang, C. M. Hu, Y. P. Chen,
and J. Q. Xiao, Cavity Mediated Manipulation of Distant Spin Currents Using a Cavity-Magnon-Polariton, Phys. Rev. Lett. {\bf 118}, 217201 (2017).

\bibitem{Mukhopadhyay-2022}D. Mukhopadhyay, J. M. P. Nair, and G. S. Agarwal, Quantum amplification of spin currents in cavity magnonics by a parametric drive induced long-lived mode, Phys. Rev. B {\bf 106}, 184426 (2022).

\bibitem{Yuanhy-2020}H. Y. Yuan, S. Zheng, Z. Ficek, Q. Y. He, and M.-H. Yung, Enhancement of magnon-magnon entanglement inside a cavity, Phys. Rev. B {\bf 101}, 014419 (2020).

\bibitem{Mousolou-2021}V. A. Mousolou, Y. Liu, A. Bergman, A. Delin, O. Eriksson, M. Pereiro, D. Thonig, and E. Sj\"{o}qvist, Phys. Magnon-magnon entanglement and its quantification via a microwave cavity, Phys. Rev. B {\bf 104}, 224302 (2021).

\bibitem{ZhangZ-2019}Z. Zhang, Marlan O. Scully, and Girish S. Agarwal, Quantum entanglement between two magnon modes via Kerr nonlinearity driven far from equilibrium, Phys. Rev. Research {\bf 1}, 023021 (2019).

\bibitem{Ren-2022}Y. Ren, J. Xie, X. Li, S. Ma, and F. Li, Long-range generation of a magnon-magnon entangled state, Phys. Rev. B {\bf 105}, 094422 (2022).

\bibitem{Pengjx2025}J.-X. Peng, S. K. Singh, N. Akhtar, Z. Gu, C. Wu, and J.-F. Li, Symmetric vs asymmetric magnon pairing: Comparative study of quantum entanglement and synchronization in the cavity-magnon system, Phys. Rev. A {\rm 112}, 042430 (2025).

\bibitem{Ahmed-2025}R. Ahmed, H. Ali, A. Shehzad, S. K. Singh, A. Sohail, and M. C. de Oliveira, Nonreciprocal multipartite entanglement induced by Kerr nonlinearity, Quantum Information Processing {\bf 24}, 137 (2025).

\bibitem{Harder-2018} M. Harder, Y. Yang, B. M. Yao, C. H. Yu, J. W. Rao, Y. S.
Gui, R. L. Stamps, and C. M. Hu, Level Attraction Due to Dissipative Magnon-Photon Coupling, Phys. Rev. Lett. {\bf 121}, 137203 (2018).

\bibitem{Grigoryan-2018}V. L. Grigoryan, K. Shen, and K. Xia, Synchronized spinphoton coupling in a microwave cavity, Phys. Rev. B. {\bf 98}, 024406 (2018).

\bibitem{Wang-2019}Y. P. Wang, J. W. Rao, Y. Yang, P. C. Xu, Y. S. Gui, B. M. Yao,
J. Q. You, and C.-M. Hu, Nonreciprocity and Unidirectional Invisibility in Cavity Magnonics, Phys. Rev. Lett. {\bf 123}, 127202 (2019).

\bibitem{Huang-2018}R. Huang, A. Miranowicz, J. Q. Liao, F. Nori, and H. Jing, Nonreciprocal Photon Blockade, Phys. Rev. Lett. {\bf 121}, 153601 (2018).

\bibitem{Yao-2022}X. Y. Yao, H. Ali, and P. B. Li, Nonreciprocal Phonon Blockade in a Spinning Acoustic Ring Cavity Coupled to a Two-Level System, Phys. Rev. Applied {\bf 17}, 054004 (2022).

\bibitem{Wangy-2022}Y. Wang, W. Xiong, Z. Xu, G. Q. Zhang, and J. Q.
You, Dissipation-induced nonreciprocal magnon blockade in a magnon-based hybrid system, Sci. China Phys. Mech. Astron. {\bf 65}, 260314 (2022).

\bibitem{Xie-2020}J. K. Xie, S. L. Ma, and F. L. Li, Quantum-interference enhanced
magnon blockade in an yttrium-iron-garnet sphere coupled to superconducting circuits, Phys. Rev. A {\bf 101}, 042331 (2020).

\bibitem{Wang-2022}F. Wang, C. Gou, J. Xu, and C. Gong, Hybrid magnon-atom entanglement and magnon blockade via quantum interference, Phys. Rev. A {\bf 106}, 013705 (2022).

\bibitem{ZhangD-2017}D. Zhang, X. Q. Luo, Y. P. Wang, T. F. Li, and J. Q. You, Observation of the exceptional point in cavity magnon-polaritons, Nat. Commun. {\bf 8}, 1368 (2017).

\bibitem{Harder-2017}M. Harder, L. Bai, P. Hyde, and C. M. Hu, Topological properties of a coupled spin-photon system induced by damping, Phys. Rev. B {\bf 95}, 214411 (2017).

\bibitem{Cao-2019}Y. Cao and P. Yan, Exceptional magnetic sensitivity of PT-symmetric cavity magnon polaritons, Phys. Rev. B {\bf 99}, 214415 (2019).

\bibitem{Zhao-2020}J. Zhao, Y. Liu, L. Wu, C. K. Duan, Y. Liu, and J. Du, Observation of Anti-PT-Symmetry Phase Transition in the Magnon-Cavity-Magnon Coupled System, Phys. Rev. Appl. {\bf 13}, 014053 (2020).

\bibitem{Zhanggq-2019}G.-Q. Zhang and J. Q. You, Higher-order exceptional point in a cavity magnonics system, Phys. Rev. B {\bf 99}, 054404 (2019).

\bibitem{Yao-2017}B. Yao, Y. S. Gui, J. W. Rao, S. Kaur, X. S. Chen, W. Lu, Y. Xiao, H. Guo, K. P. Marzlin, and C. M. Hu, Cooperative polariton dynamics in feedback-coupled cavities, Nat. Commun.	{\bf 8}, 1437 (2017).

\bibitem{Tian-2023}M. Tian, M. Wang, G.-Q. Zhang, H.-C. Li, and W. Xiong, Critical cavity-magnon polariton mediated strong long-distance spin-spin coupling, 	arXiv:2304.13553.

\bibitem{Hei-2021}X. L. Hei, X. L. Dong, J. Q. Chen, C. P. Shen, Y. F. Qiao, and
P. B. Li, Enhancing spin-photon coupling with a micromagnet, Phys. Rev. A {\bf 103}, 043706 (2021).

\bibitem{Xu-2023}D. Xu, X.-K. Gu, H.-K. Li, Y.-C. Weng, Y.-P. Wang, J. Li, H. Wang, S.-Y. Zhu, and J. Q. You, Quantum Control of a Single Magnon in a Macroscopic Spin System, Phys. Rev. Lett. {\bf 130}, 193603 (2023).

\bibitem{Yuan-2020}H. Y. Yuan, P. Yan, S. Zheng, Q. Y. He, K. Xia, and M.-H. Yung,
Steady Bell State Generation via Magnon-Photon Coupling, Phys. Rev. Lett. {\bf 124}, 053602 (2020).

\bibitem{Sun-2021}F. X. Sun, S. S. Zheng, Y. Xiao, Q. Gong, Q. He, and K.
Xia, Remote Generation of Magnon Schr\"{o}dinger Cat State via
Magnon-Photon Entanglement, Phys. Rev. Lett. {\bf 127}, 087203 (2021).

\bibitem{Zhanggq-2023}G. Q. Zhang, W. Feng, W. Xiong, Q. P. Su, and C. P. Yang,
Generation of long-lived W states via reservoir engineering in dissipatively coupled systems, Phys. Rev. A {\bf 107}, 012410 (2023).

\bibitem{Qi-2022}S. F. Qi and J. Jing, Generation of Bell and Greenberger-Horne-Zeilinger states from a hybrid qubit-photon-magnon system,
Phys. Rev. A {\bf 105}, 022624 (2022).

\bibitem{Hisatomi-2016}R. Hisatomi, A. Osada, Y. Tabuchi, T. Ishikawa, A. Noguchi, R. Yamazaki, K. Usami, and Y. Nakamura, Bidirectional conversion between microwave and light via ferromagnetic magnons, Phys. Rev. B {\bf 93}, 174427 (2016).

\bibitem{Zhu-2020}N. Zhu, X. Zhang, X. Han, C. L. Zou, C. Zhong, C. H. Wang,
L. Jiang, and H. X. Tang, Waveguide cavity optomagnonics for broadband multimode microwave-to-optics conversion, Optica {\bf 7}, 1291 (2020).

\bibitem{Tabuchi-2015}Y. Tabuchi, S. Ishino, A. Noguchi, T. Ishikawa, R. Yamazaki, K. Usami, and Y. Nakamura, Coherent coupling between a ferromagnetic magnon and a superconducting qubit, Science {\bf 349}, 405 (2015).

\bibitem{Lachance-2020}D. Lachance-Quirion, S. P. Wolski, Y. Tabuchi,
S. Kono, K. Usami, and Y. Nakamura, Entanglement-based single-shot detection of a single magnon with a superconducting qubit, Science, {\bf 367}, 425 (2020).

\bibitem{Dobrovolskiy-2019}O. V. Dobrovolskiy, R. Sachser, T. Br\"{a}cher, T. B\"{o}ttcher, V. V. Kruglyak, R. V. Vovk, V. A. Shklovskij, M. Huth, B. Hillebrands, and A. V. Chumak, Magnon–fluxon interaction in a ferromagnet/superconductor heterostructure, Nat. Phys. {\bf 15}, 477 (2019). 

\bibitem{Wolski-2020}S. P. Wolski, D. Lachance-Quirion, Y. Tabuchi, S. Kono, A. Noguchi, K. Usami, and Y. Nakamura, Dissipation-Based Quantum Sensing of Magnons with a Superconducting Qubit, Phys. Rev. Lett. {\bf 125}, 117701 (2020). 

\bibitem{Xiong4-2022}W. Xiong, M. Tian, G.-Q. Zhang, and J. Q. You, Strong long-range spin-spin coupling via a Kerr magnon interface, Physical Review B {\bf105}, 245310 (2022).


\bibitem{neuman-2020}T. Neuman, D. S. Wang, and P. Narang, Nanomagnonic Cavities for Strong Spin-Magnon Coupling and Magnon-Mediated Spin-Spin Interactions, Phys. Rev. Lett. {\bf 125}, 247702 (2020).

\bibitem{neuman1-2021} D. S. Wang, T. Neuman, and P. Narang, Spin Emitters beyond the Point Dipole Approximation in Nanomagnonic Cavities, J. Phys. Chem. C {\bf 125}, 6222 (2021).

\bibitem{neuman2-2021}D. S. Wang, M. Haas, and P. Narang, Quantum Interfaces to the Nanoscale, ACS Nano {\bf 15}, 7879 (2021).

\bibitem{Skogvoll-2021}I. C. Skogvoll, J. Lidal, J. Danon, and A. Kamra, Tunable anisotropic quantum Rabi model via magnon spin-qubit ensemble, Phys. Rev. Applied {\bf 16}, 064008 (2021).

\bibitem{trifunovic-2013}L. Trifunovic, F. L. Pedrocchi, and D. Loss, Long-Distance Entanglement of Spin Qubits via Ferromagnet, Phys. Rev. X {\bf 3}, 041023 (2013).

\bibitem{Fukami-2021}M. Fukami, D. R. Candido, D. D. Awschalom, and M. E. Flatt$\acute{\rm  e}$, Opportunities for Long-Range Magnon-Mediated Entanglement of Spin Qubits via On- and Off-Resonant Coupling, PRX Quantum {\bf 2}, 040314 (2021).

\bibitem{zhang-2016}X. Zhang, C. L. Zou, L. Jiang, and H. X. Tang, Cavity magnomechanics, Sci. Adv. {\bf 2}, e1501286 (2016).

\bibitem{Li-2018}J. Li, S.-Y. Zhu, and G. S. Agarwal, Magnon-Photon-Phonon Entanglement in Cavity Magnomechanics, Phys. Rev. Lett. {\bf 121}, 203601 (2018).

\bibitem{Shen-2022}R.-C. Shen, J. Li, Z.-Y. Fan, Y.-P. Wang, and J. Q. You, Mechanical Bistability in Kerr-modified Cavity Magnomechanics, Phys. Rev. Lett. 129, 123601 (2022).

\bibitem{Li-2021}J. Li, Y.-P. Wang, W.-J. Wu, S.-Y. Zhu, and J. Q. You, Quantum Network with Magnonic and Mechanical Nodes, PRX Quantum {\bf 2}, 040344 (2021).

\bibitem{Sohail-2025}A. Sohail, M. Amazioug, S. K. Singh, N. Chabar, R. Ahmed, M. C. de Oliveira, Coherent Feedback Control of Indirectly Coupled Mode Multipartite Entanglement in a Cavity Opto-Magnomechanical System, Annalen der Physik {\bf 537}, 2400375 (2025).

\bibitem{Faisal-2025}F. H. A. Mathkoor, S. K. Singh, R. Ahmed, J.-X. Peng, M. Amazioug, M. Khalid, and A. Sohail, Bipartite entanglement and Gaussian quantum steering in a whispering gallery mode coupled with two magnon modes, Scientific Reports {\bf 15}, 13503 (2025).

\bibitem{Sohail-2023}A. Sohail, J.-X. Peng, A. Hidki, M. Khalid, and S. K. Singh, Distant entanglement via photon hopping in a coupled cavity magnomechanical system, Scientific Reports {\bf 13}, 21840 (2023).

\bibitem{Sohail-2023b}A. Sohail, R. Ahmed, J.-X. Peng, A. Shahzad, and S. K. Singh, Enhanced entanglement via magnon squeezing in a two-cavity magnomechanical system, Journal of the Optical Society of America B {\bf 40}, 1359 (2023).

\bibitem{Chen}J. Chen, X.-G. Fan, W. Xiong, D. Wang, and L. Ye, Nonreciprocal entanglement in cavity-magnon optomechanics, Phys. Rev. B {\bf 108}, 024105 (2023).

\bibitem{Proskurin-2018}I. Proskurin, A. S. Ovchinnikov, J. Kishine, and R. L. Stamps, Cavity optomechanics of topological spin textures in magnetic insulators, Phys. Rev. B {\bf 98}, 220411(R) (2018).

\bibitem{Xiong-2023}W. Xiong, M. Wang, G.-Q. Zhang, and J. Chen, Optomechanical-interface-induced strong spin-magnon coupling, Phys. Rev. A {\bf 107}, 033516 (2023).

\bibitem{Berinyuy2025}E. K. Berinyuy, J.-X. Peng, A. Sohail, P. Djorwé, A.-H. Abdel-Aty, N. Alessa, K.S. Nisar, and S. G. Nana Engo,  Nonreciprocal entanglement in a molecular optomechanical system, Physica B: Condensed Matter {\bf713}, 417313 (2025).

\bibitem{Sohail2023}A. Sohail, M. Qasymeh, and H. Eleuch, Entanglement and quantum steering in a hybrid quadpartite system, Phys. Rev. Applied {\bf 20}, 054062 (2023)

\bibitem{Gao-2017}Y.-P. Gao, C. Cao, T.-J. Wang, Y. Zhang, and C. Wang, Cavity-mediated coupling of phonons and magnons, Phys. Rev. A {\bf 96}, 023826 (2017).

\bibitem{Zhangx-2016}X. Zhang, N. Zhu, C.-L. Zou, and H. X. Tang, Optomagnonic Whispering Gallery Microresonators, Phys. Rev. Lett. {\bf 117}, 123605 (2016).

\bibitem{Osada-2016}A. Osada, R. Hisatomi, A. Noguchi, Y. Tabuchi, R. Yamazaki, K. Usami, M. Sadgrove, R. Yalla, M. Nomura, and Y. Nakamura, Cavity Optomagnonics with Spin-Orbit Coupled Photons, Phys. Rev. Lett. {\bf 116}, 223601 (2016).

\bibitem{Haigh-2016}J. A. Haigh, A. Nunnenkamp, A. J. Ramsay, and A. J. Ferguson, Triple-Resonant Brillouin Light Scattering in Magneto-Optical Cavities, Phys. Rev. Lett. {\bf 117}, 133602 (2016).

\bibitem{Zhangguoq-2019}G. Q. Zhang, Y. P. Wang, and J. Q. You, Theory of the magnon Kerr effect in cavity magnonics, Sci. China: Phys. Mech. Astron. {\bf 62}, 987511 (2019).

\bibitem{Wangyp-2018}Y. P. Wang, G. Q. Zhang, D. Zhang, T. F. Li, C. M. Hu, and J. Q. You, Bistability of Cavity Magnon Polaritons, Phys. Rev. Lett. {\bf 120}, 057202 (2018).

\bibitem{Zheng-2023}S. Zheng, Z. Wang, Y. Wang, F. Sun, Q. He, P. Yan, and H. Y. Yuan, Tutorial: Nonlinear magnonics, arXiv:2303.16313.

\bibitem{Shenrc-2021} R. C. Shen, Y. P. Wang, J. Li, S. Y. Zhu, G. S. Agarwal, and J. Q. You, Long-Time Memory and Ternary Logic Gate Using a Multistable Cavity Magnonic System, Phys. Rev. Lett. {\bf 127}, 183202 (2021).

\bibitem{Ji-2023}F.-Z. Ji and J.-H. An, Kerr-Nonlinearity-Induced Strong Spin-Magnon Coupling, arXiv:2308.05927.

\bibitem{Peng2025}M. L. Peng, M. Tian, X. C. Chen, M. F. Wang, G. Q. Zhang, H. C. Li, and W. Xiong, Cavity magnon-polariton interface for strong spin-spin coupling, Opt. Lett. {\bf 50}, 1516 (2025).

\bibitem{Liu-2023}G. Liu, W. Xiong, and Z. J. Ying, Switchable superradiant phase transition with Kerr magnons, Phys. Rev. A {\bf 108}, 033704 (2023).

\bibitem{Zhanggq1-2023} G.-Q. Zhang, Y. Wang, and W. Xiong, Detection sensitivity enhancement of magnon Kerr nonlinearity in cavity magnonics induced by coherent perfect absorption, Phys. Rev. B {\bf 107}, 064417 (2023).

\bibitem{Nair-2021}J. M. P. Nair, D. Mukhopadhyay, and G. S. Agarwal, Enhanced sensing of weak anharmonicities through coherences in dissipatively coupled anti-PT symmetric systems, Phys. Rev. Lett. {\bf 126}, 180401 (2021).

\bibitem{Chen2024}J. Chen, X.-G. Fan, W. Xiong, D. Wang, and L. Ye, Nonreciprocal Photon-Phonon Entanglement in Kerr-Modified Spinning Cavity Magnomechanics, Phys. Rev. A {\bf 109}, 043512 (2024).

\bibitem{Liu2025}M. Y. Liu, Y. Gong, J. Chen, Y. W. Wang, and W. Xiong, Nonreciprocal Microwave-Optical Entanglement in Kerr-Modified Cavity Optomagnomechanics, Chin. Phys. B {\bf 34}, 057202 (2025).

\bibitem{Kong-2019}C. Kong, H. Xiong, and Y. Wu, Magnon-Induced Nonreciprocity Based on the Magnon Kerr Effect, Phys. Rev. Applied {\bf 12}, 034001 (2019).

\bibitem{Fan2024}X. H. Fan, Y. N. Zhang, J. P. Yu, M. Y. Liu, W. D. He, H. C. Li, and W. Xiong, Nonreciprocal Unconventional Photon Blockade with Kerr Magnons, Adv. Quan. Tech. {\bf 7}, 2400043 (2024).

\bibitem{Lai2025}D. G. Lai, A. Miranowicz, and F. Nori, Nonreciprocal quantum synchronization, Nat. Commun. 16, 8491 (2025).

\bibitem{Wangm-2021}M. Wang, C. Kong, Z.-Y. Sun, D. Zhang, Y.-Y. Wu, and L.-L. Zheng, Nonreciprocal high-order sidebands induced by magnon Kerr nonlinearity, Phys. Rev. A {\bf 104}, 033708 (2021).

\bibitem{Lorentz-1896}H. A. Lorentz, Eene algemeene stelling omtrent de beweging eener vloeistof met wrijving en eenige daaruit afgeleide gevolgen, Versl. Kon. Acad. Wet. Amst. {\bf 5}, 168 (1896).

\bibitem{Masoud-2019} H. Masoud and H. A. Stone, The reciprocal theorem in fluid dynamics and transport phenomena, J. Fluid Mech. {\bf 879}, 1 (2019).

{\bibitem{Qiu-2023}J. Y. Qiu, A. Grimsmo, K. Peng, B. Kannan, B. Lienhard, Y. Sung, P. Krantz, V. Bolkhovsky, G. Calusine, D. Kim, and A. Melville, Broadband squeezed microwaves and amplification with a Josephson travelling-wave parametric amplifier, Nat. Phys. {\bf 19}, 706 (2023).}

{\bibitem{Mahboob-2022}I. Mahboob, H.  Toida,  K.  Kakuyanagi, Y. Nakamura, and S. Saito, A three-dimensional Josephson parametric amplifier, Appl. Phys. Exp. {\bf 15}, 062005 (2022).}

\bibitem{Yurke-1987}B. Yurke, Squeezed-state generation using a Josephson parametric amplifier, Journal of the Optical Society of America B {\bf 4}, 1551 (1987).

\bibitem{Yurke-1988} B. Yurke, P. Kaminsky, R. Miller, E. Whittaker, A. Smith, A. Silver, and R. Simon, Observation of 4.2-K equilibrium-noise squeezing via a Josephson-parametric amplifier, Phys. Rev. Lett.
{\bf 60}, 764 (1988).

\bibitem{Yurke-1989}B. Yurke, L. Corruccini, P. Kaminsky, L. Rupp, A. Smith, A. Silver, R. Simon, and E. Whittaker, Observation of parametric amplification and deamplification in a Josephson parametric amplifier, Phys. Rev. A {\bf 39}, 2519 (1989).

\bibitem{Movshovich1990}R. Movshovich, B. Yurke, P. Kaminsky, A. Smith, A. Silver, R. Simon, and M. Schneider, Observation of zero-point noise squeezing via a Josephson-parametric amplifier, Phys. Rev. Lett. {\bf 65}, 1419 (1990).

\bibitem{Castellanos-Beltran2008}M. A. Castellanos-Beltran, K. Irwin, G. Hilton, L. Vale, and K. Lehnert, Amplification and squeezing of quantum noise with a tunable Josephson metamaterial, Nat. Phys. {\bf 4}, 929 (2008).

\bibitem{Yamamoto}T. Yamamoto, K. Inomata, M. Watanabe, K. Matsuba, T. Miyazaki, W. D. Oliver, Y. Nakamura, and J. Tsai, Flux-driven Josephson parametric amplifier,  Flux-driven Josephson parametric amplifier, Appl. Phys. Lett. {\bf 93}, 042510 (2008).

\bibitem{Mallet2011}F. Mallet, M. Castellanos-Beltran, H. Ku, S. Glancy,
E. Knill, K. Irwin, G. Hilton, L. Vale, and K. Lehnert, Quantum state tomography of an itinerant squeezed microwave field, 
Phys. Rev. Lett. {\bf 106}, 220502 (2011).

\bibitem{Fedorov2016}K. G. Fedorov, L. Zhong, S. Pogorzalek, P. Eder, M. Fischer, J. Goetz, E. Xie, F. Wulschner, K. Inomata, T. Yamamoto, T. Yamamoto, Y. Nakamura, R. Di Candia, U. Las Heras, M. Sanz, E. Solano, E. P. Menzel, F. Deppe, A. Marx, and R. Gross, Displacement of propagating squeezed microwave states, Phys. Rev. Lett. {\bf 117}, 020502 (2016).

\bibitem{Kono2017}S. Kono, Y. Masuyama, T. Ishikawa, Y. Tabuchi, R. Yamazaki, K. Usami, K. Koshino, and Y. Nakamura, Nonclassical photon number distribution in a superconducting cavity under a squeezed drive, Phys.
Rev. Lett. {\bf 119}, 023602 (2017).

\bibitem{Bienfait2017}A. Bienfait, P. Campagne-Ibarcq, A. Kiilerich, X. Zhou, S. Probst, J. Pla, T. Schenkel, D. Vion, D. Estève, J. Morton, J. J. L. Morton, K. Moelmer, and P. Bertet, Magnetic Resonance with Squeezed Microwaves, Phys. Rev. X {\bf 7}, 041011 (2017).



\end{thebibliography}

\end{document}